\documentstyle[multicol,prl,aps,psfig,floats]{revtex}

\begin{document}

\draft

\title{Statistics of Shear-induced Rearrangements in a Model Foam}
\author{Shubha Tewari$^{1,2}$, Dylan Schiemann$^{2}$, Douglas J.
Durian$^{1}$, Charles
M. Knobler$^{2}$\\
Stephen A. Langer$^{3}$ and Andrea J. Liu$^{2}$}
\address{$^{1}$UCLA Department of Physics and Astronomy\\
$^{2}$UCLA Department of Chemistry and Biochemistry\\
Los Angeles, CA 90095\\
$^{3}$Information Technology Laboratory, NIST\\
Gaithersburg, MD 20899}

\date{\today }

\maketitle

\begin{abstract}
Under steady shear, a foam relaxes stress through intermittent
rearrangements of bubbles accompanied by sudden drops in the
stored elastic energy.  We use a simple model of foam that incorporates both
elasticity and dissipation to study the
statistics of bubble rearrangements in terms of energy drops,
the number of nearest neighbor changes, and the rate of
neighbor-switching (T1) events.  We do this for a two-dimensional
system as a function of system size, shear rate, dissipation
mechanism, and gas area fraction.  We find that for dry foams, there is a
well-defined quasistatic limit at low shear rates where localized
rearrangements occur at a constant rate per unit strain, independent
of both system size and dissipation mechanism.  These results are in good
qualitative agreement with experiments on two-dimensional and
three-dimensional foams. In contrast, we find for progessively wetter foams
that the event size distribution broadens
into a power law that is cut off only by system size.  This is consistent with
criticality at the melting transition.
\end{abstract}


\begin{multicols}{2}

\section{Introduction}

A foam is a disordered collection of densely-packed polydisperse gas
bubbles in a relatively small volume of liquid
\cite{prudhommebook,djddaw,weaire84}.
Foams have a rich rheological behavior; they act like elastic solids
for small deformations but they flow like viscous liquids at large
applied shear stress \cite{kranikrev}.  The stress is relaxed by
discrete rearrangement events that occur intermittently as the foam is
sheared.  Three-dimensional foams are opaque, which makes it difficult
to observe these bubble movements directly.  However, measurements
\cite{gopal,gopal2} by diffusing-wave spectroscopy of
three-dimensional foams subjected to a constant shear rate suggest
that the number of bubbles involved in the rearrangements is small, of
the order of four bubbles.  Bubble rearrangements can be observed
directly by fluorescence microscopy in two-dimensional foams
found in insoluble
monolayers at the air-water interface.  A study of shear in such
foams \cite{dennin}
also revealed no large-scale rearrangements.

While analytical theories for the response to applied steady shear may be
constructed for periodic foams, only simulation approaches are
possible for disordered foams.  Kawasaki's \cite{okuzono} vertex model
was the first to incorporate dissipative dynamics.  It applies to a
two-dimensional foam in the limit in which the area fraction of gas is
unity (a dry foam).  Bubble edges are approximated by straight line
segments that meet at a vertex that represents a Plateau border.  The
equations of motion for the vertices are solved by balancing
viscous dissipation due to shear flow within the borders by surface
tension forces.  At low shear rates, the elastic energy of the foam,
which is associated with the total length of the bubble segments,
shows intermittent energy drops with a distribution of event rate vs.
energy release that follows a broad power law, consistent with
self-organized criticality.  The rearrangements associated with the
largest events consist of cooperative motions of bubbles that extend
over much of the system.

Weaire and coworkers \cite{bolton0,bolton,hutzler} were the first to
develop a model appropriate to a disordered wet foam.  The
model does not include dissipation, so it is quasi-static by
construction.  Thus the system is allowed to relax to an equilibrium
configuration after each of a series of infinitesimal shear steps.
The size of rearrangements is measured by the number of changes in
nearest-neighbor contacts.  For dry foams, the average event size is
small, inconsistent with a picture of self-organized criticality.
However, as the liquid content increases, the event-size distribution
broadens, with the largest events involving many bubbles.  Although the
statistics are limited, this is consistent with a picture of
criticality at the point where the foam loses its rigidity.

The first model capable of treating wet, disordered foams at nonzero
shear rate was proposed by Durian \cite{durian1}.  His model pictures
the foam as consisting of spherical bubbles that can overlap.  Two
pairwise-additive interactions between neighboring bubbles are
considered, a harmonic repulsive force that mimics the effect of
bubble deformation and a force proportional to the velocity difference
between neighboring bubbles that accounts for the viscous drag.  He
found \cite{durian2} that the probability density of energy drops
followed a power law, with a cutoff at very high energy
events.  The
largest event observed consisted of only a few bubbles changing
neighbors.  This is inconsistent with a picture of self-organized
criticality, although the effect of the liquid content on the topology
statistics was not
examined.

Most recently, Jiang et al.\cite{jiang} have employed a large-Q
Potts model to examine sheared foams.  In this lattice model bubbles
are represented by domains of like spin, and the film boundaries are
the links between regions of different spins.  Each spin merely acts
as a label for a particular bubble, and the surface energy arises only
at the boundaries where the spins differ.  The evolution of the foam
is studied by Monte Carlo dynamics with a Hamiltonian consisting of
three terms: the coupling energy between neighboring spins at the
boundaries of the bubbles; an energy penalty for changes in the areas
of the bubbles, which inhibits coarsening of the foam; and a shear
term that biases the probability of a spin reassignment in the strain
direction.  The spatial distribution of T1 events was examined and no
system-wide rearrangements were observed.  Nevertheless, Jiang, et al.
found a power-law distribution of energy changes.  They also found
that the number of events per unit strain displayed a strong shear
rate dependence, suggesting that a quasi-static limit does not exist.

These four simulation approaches thus offer conflicting pictures as to
(1) the existence of a quasistatic limit, (2) whether or not
rearrangement dynamics at low shear rates are a form of self-organized
criticality, and (3) whether or not the melting of foams with
increasing liquid content is a more usual form of criticality.  One
possible reason for this disagreement is differences in the treatment of
dissipation, and hence in the treatment of the {\it dynamics} of the
rearrangements.  In principle, the only accurate way in which to
include dissipation in a sheared foam is to solve for the Stokes flow
in the liquid films and Plateau borders.  This approach has been
adopted by Li, Zhou and Pozrikidis\cite{pozrikidis}, but so far it has only
been applied to periodic foams.  The statistics of rearrangement
events are fundamentally different in periodic and disordered foams;
in sheared periodic foams, all the bubbles rearrange simultaneously at
periodic intervals, while in a disordered foam, the rearrangements can
be localized and intermittent.  Nonetheless, the Stokes-flow approach
is the only one that can be used as a benchmark for more simplified
models.

In order to gain a better understanding of the origin of the
discrepancies between the various models, as well as between the
models and experiments, we report here a systematic study of the
properties of a sheared foam using Durian's model.  We begin by
reviewing his model and discussing our numerical implementation using
two different forms of dissipation.  After confirming that there are
no significant system-size effects for dry samples, we examine
shear-rate dependence and establish the existence of a true
quasistatic limit for the distribution and rate of energy drops and
topology changes.  This limit is shown to be independent of the dissipation
mechanism for foams of different gas fractions.  Finally,
we examine dramatic changes in the behavior of these quantities as the
liquid content is tuned toward the melting point.

\section{Bubble model}

Durian's model\cite{durian1,durian2} is based on
the wet-foam limit, where the bubbles are spherical.  The foam is
described entirely in terms of the bubble radii $\{ R_i \}$ and the
time-dependent positions of the bubble centers $\{ \vec{r}_i \}$.  The
details of the microscopic interactions at the level of soap films and
vertices are subsumed into two pairwise additive interactions between
bubbles, which arise when the distance between bubble centers is less
than the sum of their radii.  The first, a repulsion that originates
in the energy cost to distort bubbles, is modeled by the compression
of two springs in series with individual spring constants that scale
with the Laplace pressures $\sigma/R_i$, where $\sigma$ is the
liquid-gas surface tension and $R_i $ is the bubble radius.  Bubbles
that do not overlap are assumed not to interact.  The repulsive force
on bubble $i$ due to bubble $j$ is then
\begin{equation}
\vec{F}_{ij}^r = k_{ij}  \left [(R_i + R_j) -
|\vec{r}_i - \vec{r}_j| \right ] \hat{r}_{ij}
\end{equation}
where $\hat{r}_{ij}$ is the unit vector pointing from the center of
bubble $j$ to the center of bubble $i$, and $k_{ij} = F_0/(R_i +R_j)$ is the
effective spring constant, with $F_0 \approx \sigma \langle R
\rangle$.  The second interaction is the viscous dissipation due to
the flow of liquid in the films.  It, too, is assumed to be pairwise
additive and is modeled by the simplest form of drag, where the force
is proportional to the velocity difference between overlapping
bubbles.  The viscous force on bubble $i$ due to its neighbor $j$ is
\begin{equation}
\vec{F}_{ij}^v = -b(\vec{v}_i - \vec{v}_j),
\label{drag}
\end{equation}
where the constant $b$ is proportional to the viscosity of the liquid,
and is assumed to be the same for all bubble neighbors.

The net force on each bubble sums to zero, since inertial effects are
negligible in this system.  Summing over those bubbles $j$ that
touch bubble $i$, the equation of motion for bubble $i$ is
\begin{equation}
\sum_j (\vec{v}_i - \vec{v}_j) = \frac{F_0}{b} \sum_j \left [
\frac{1}{|\vec{r}_i - \vec{r}_j|} - \frac{1}{R_i + R_j} \right ]
(\vec{r}_i - \vec{r}_j) + \frac{\vec{F}_i^a}{b},
\label{eq:vel}
\end{equation}
where $\vec{F}_i^a$ is an externally applied force, arising, for
instance, from interactions with moving walls.

Durian \cite{durian1,durian2} employed a further simplification of this model,
in
which the viscous dissipation is taken into account in a mean-field
manner by taking the velocity of each bubble relative to an average
linear shear profile.  In this case, the total drag force on bubble $i$
due to all of its $N_{i}$ overlapping neighbors is
\begin{equation}
\vec{F}_{i}^v = -b N_{i} \left ( \vec{v}_i - \dot{\gamma}y_i \hat{x}\right ).
\label{mf}
\end{equation}
In the numerical simulations reported here we use both the mean-field
model of dissipation as well as the approximation represented by
Eq.~\ref{drag}, which we call the local dissipation model.  In the
latter, at each integration time step the velocity of a bubble is
measured with respect to the average of the velocities of its $N_i$
overlapping neighbors, so that the total drag force on bubble $i$ is
\begin{equation}
\vec{F}_{i}^v = -b \left (N_{i} \vec{v}_i - \sum_{j={\text{
nn}}} \vec{v}_j \right )
\label{local}
\end{equation}
For very large $N_{i}$, this reduces to Eq.~\ref{mf}; otherwise, it
allows for fluctuations.
One aim of our study is to establish the sensitivity of the results to
the specific form of dissipation used, Eq.~\ref{mf} or
Eq.~\ref{local}.

In two dimensions, the area fraction of gas bubbles,
$\phi$, can be defined by the total bubble area $\sum\pi R_{i}^{2}$
per system area.
Because the bubbles are constrained to remain circular and their
interactions are approximated as pairwise-additive\cite{lacasse},
the model necessarily breaks down for very dry foams.  In fact, bubble radii
can even be chosen so that $\phi$ exceeds one.  In a real foam, of
course, this is prevented by the divergence of the osmotic pressure.

\section{Numerical Method}

All the results reported here are based on simulations of a
two-dimensional version of Durian's model.  We use Eq.~\ref{eq:vel}
to study a two-dimensional foam periodic in the $x$--direction and
trapped between parallel plates in the $y$--direction.  Bubbles that
touch the top and bottom plates are fixed to them, and the top plate
is moved at a constant velocity in the $x$--direction.  (The system
can also be sheared with a constant force instead of a constant
velocity, but that case will not be discussed here.)  Thus, bubbles
are divided into two categories --- ``boundary'' bubbles, which have
velocities that are determined by the motion of the plates, and
``interior'' bubbles, whose velocities must be determined from the
equations of motion.

The equation of motion Eq.~\ref{eq:vel} can be written in the form
\begin{equation}
\label{eq:Meom}
{\bf M}(\{{\bf r}\})\cdot\{{\bf v}\} = \{{\bf F}^r\}/b + \{{\bf
F}^a\}/b
\end{equation}
where $\{\bf v\}$ is a vector containing all the velocity components
of all of the bubbles, $\{v_0^x, v_0^y, v_1^x, v_1^y, \ldots\}$,
$\{{\bf F}^r\}$ is a vector of all of the repulsive bubble--bubble
forces, and $\{{\bf F}^a\}$ contains all the forces exerted by
the walls.  The matrix $\bf M$ depends on the instantaneous positions
of the bubbles.  The $2\times2$ block submatrix $M_{ij}$ is a unit
matrix $\bf 1$ if the distinct bubbles $i$ and $j$ overlap, and $\bf
0$ if
they do not overlap.  On the diagonal, $M_{ii} = -{\bf 1} N_i$, where
$N_i$ is the number of overlapping neighbors of bubble $i$.  Eq.
\ref{eq:Meom} is of the form ${\bf A}({\bf r}, t)\cdot (d{\bf r}/dt) =
f({\bf r}, t)$, which we solve for the bubble positions $\bf r$ with
the routine DDRIV3\cite{libs}.  DDRIV3 has the ability to solve
differential equations in which the left hand side is multiplied by an
arbitrary time-dependent matrix.  Furthermore, it allows all matrix
algebra to be performed by external routines, allowing us to take
advantage of the sparse nature of $\bf M$.  We use the
SPARSKIT2\cite{libs} library for sparse matrix solutions.

The only relevant dynamical scale in this problem is set by the
characteristic relaxation time arising from the competing mechanisms
for elastic storage and viscous dissipation, $\tau_d= b\langle R
\rangle/F_0$.  This is the characteristic time scale for the duration of bubble
rearrangements driven by a drop in total elastic energy.  Without loss of
generality we set this to unity in the simulation.  In these units,
the dimensionless shear rate $\dot \gamma$ is the capillary number.

To introduce polydispersity, the bubble radii are drawn at random from
a flat distribution of variable width; in all the results reported
here, the bubble radii vary from 0.2 to 1.8 times the average bubble
radius.  We
note that the size distribution in experimental systems is closer to a
truncated Gaussian with the maximum size equal to twice the average
radius.  The truncated Gaussian distribution arises naturally from the
coarsening process
\cite{glazierweaire,stavansrev}.  We tested the sensitivity of our
results to the bubble distribution by doing one run with bubbles drawn
from a triangular distribution, and found that the shape of the
distribution had
no significant effect.  Similarly, variation of the width of a triangular
distribution has been shown to have no influence on the linear
viscoelasticity \cite{durian2}.  Note that it is important to include
polydispersity because a monodisperse system will crystallize under
shear, especially in two dimensions.

In all of our runs, the system is first equilibrated with all bubbles
treated as interior bubbles, and with a repulsive interaction between
the bubbles and the top and bottom plates so that bubbles cannot
penetrate the plates.  The bubbles that touch the top and
bottom plates are then converted to boundary bubbles.  The top plate
is moved at a constant velocity and data collection begins after any
initial transients die away.  In addition to recording quantitative
measures of the system, we also run movies of the sheared foam in
order to observe visually how the flow changes as a function of shear
rate, area fraction and other parameters \cite{movies}.

\section{Quantities Measured}

Before showing results, we discuss the various quantities extracted
during a run.  Under a small applied shear strain, bubbles in a real
foam distort; as the shear strain increases, the structure can become
unstable and they may thus rearrange their relative
positions.  In the bubble model, the distortion of bubbles is measured
globally by the total elastic energy stored in all the springs connecting
overlapping bubbles:
\begin{equation}
E = \sum {1 \over 2}k_{ij}  \left [(R_i + R_j) -
|\vec{r}_i - \vec{r}_j| \right ] ^{2}.
\label{Etot}
\end{equation}
Under steady shear, the elastic energy rises as bubbles
distort (overlap) and then drops as bubbles rearrange.  Thus, the
total elastic energy fluctuates around some average value.  The scale
of the energy is set by the elastic interaction and is of order $F_0
\langle R \rangle$ per bubble, where $ \langle R\rangle $ is the average
bubble radius.

\begin{figure}
\label{fig1}
\centerline{\psfig{file=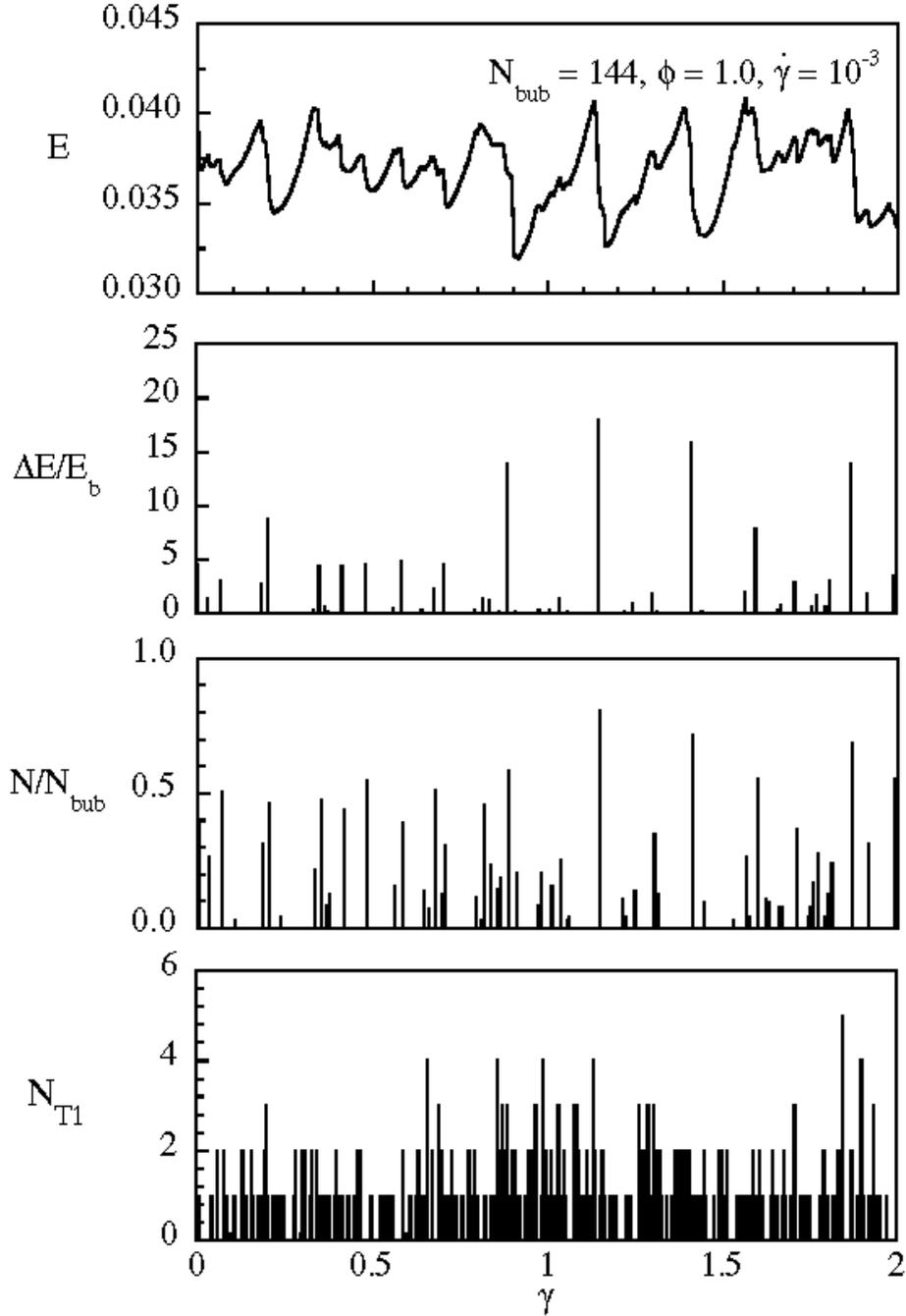,width=5in}}
\caption{Elastic energy and rearrangements vs strain for a 144-bubble system,
with gas (area) fraction $\phi=1$, being slowly sheared at
rate $\dot{\gamma}= 10^{-3}$.  The top plot (a) shows the total
elastic energy stored in the ``springs'' of overlapping bubbles.
Plot (b)  shows the size of the energy drops that occur as the
system is sheared.  Note that the duration of an energy drop is very
short compared to the time between energy drops at this low shear rate.
Plot (c) shows the corresponding fraction of bubbles that
experience a change in overlapping neighbors during each precipitous
energy-drop event.  The bottom plot (d) marks the mid-point of each T1
event, where two bubbles begin to intrude between two others; these have no
direct correspondence to the energy drop events seen in (a) and (b).
The behavior of all the properties shown here indicates that flow is
accomplished inhomogeneously and
intermittently by sudden rearrangements.}
\end{figure}

Fig.~\ref{fig1}a shows a plot of the total elastic energy as a
function of strain for a system of 144 bubbles at area fraction
$\phi=1.0$ driven at a constant shear rate of $\dot \gamma =10^{-3}$.
Similar plots for stress vs strain are shown in Refs.
\cite{durian1,durian2}.
Note the precipitous energy drops, $\Delta E$, due to bubble
rearrangements.  In the literature, these energy drops are often
referred to as avalanches.  Since the term ``avalanche'' tends to
imply the existence of self-organized criticality, we employ the more
neutral but less elegant term ``energy drop.''  The time interval
between energy drops is much larger than the duration of a single
event.  This is also illustrated in Fig.~\ref{fig1}b, which shows the
magnitude of energy drops that occur as the system is strained.  ($\Delta
E$ is normalized by the average energy per bubble $E_b$, which has been
computed by averaging the elastic energy over the entire duration of a run
and dividing by the total number of bubbles in the system, $N_{bub}$.)  These
recurring precipitous rearrangements represent the only
way for the foam to relax stress: there is no mechanism involving a
gradual energy release, as illustrated in Fig.~\ref{fig1}a.  Note
that we compute only the total elastic energy of the system; because
events can be localized and intermittent, the elastic energy may be
dropping in one region of the sample and rising in other regions.
This would limit the size of the energy drop measured.

\begin{figure}
\label{edropav}
\centerline{\psfig{file=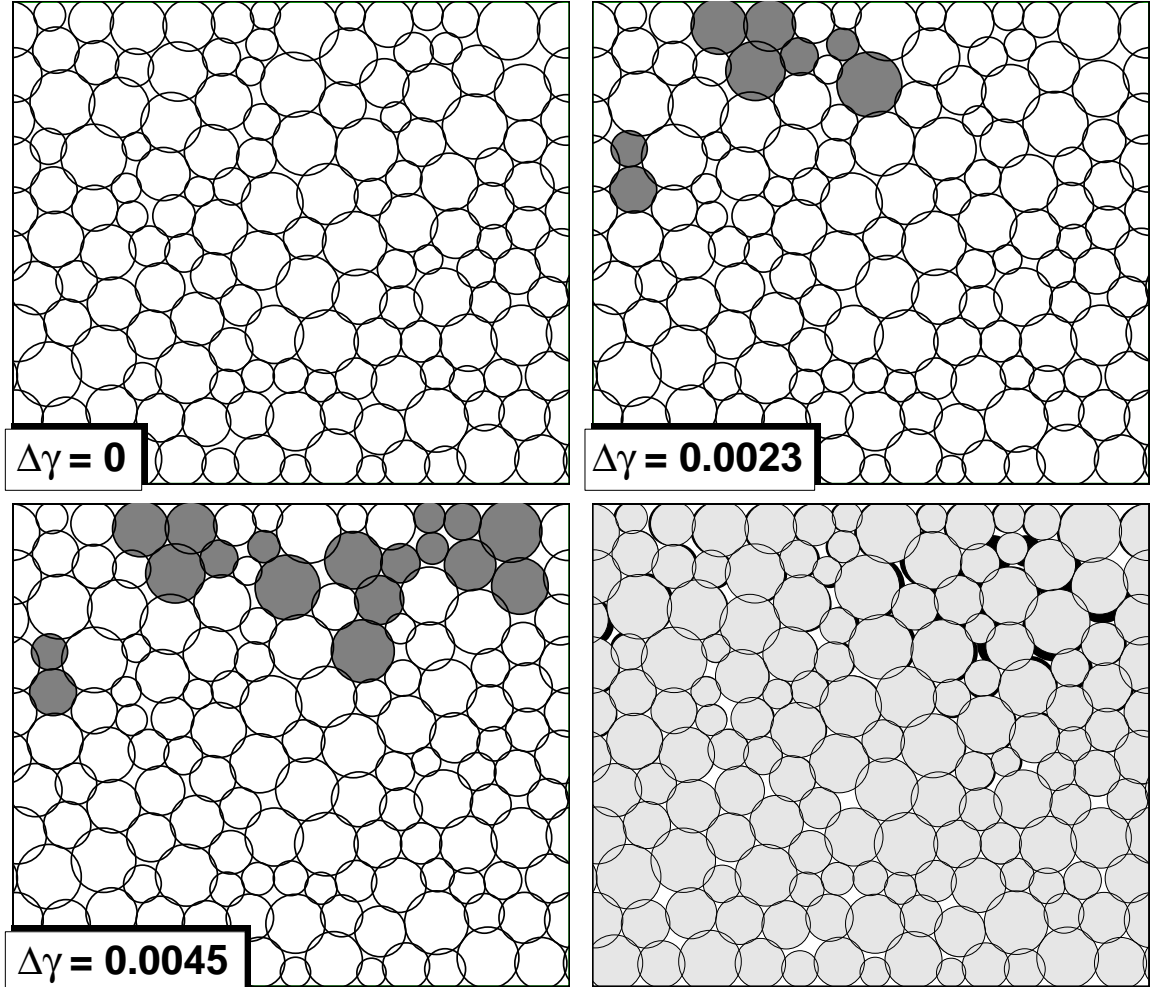,width=6in}}
\caption{Sequence of snapshots showing the nature of bubble
rearrangements during an energy drop of average size in a 144-bubble
system at $\phi=1.0$ sheared at a rate of $\dot\gamma=10^{-3}$.
The magnitude of
the drop ($\Delta E/E_{b}=2.61$) and the fraction of
bubbles that change overlapping neighbors
($N/N_{bub}=0.18$) are both close to average.  The first three frames show
the configurations of bubbles at the start, middle and end of the
energy drop, respectively.
As the event proceeds, more and more bubbles change overlapping
neighbors, as shown by the gray bubbles.  The fourth frame shows the
final configuration with bubbles in light gray superimposed on the initial
configuration with bubbles in black.  Most of the bubble motions
involve subtle shifts of bubble positions; there are no topological
rearrangements in this event.  Note that although this event appears
to nucleate at the top, in general the events appear
randomly throughout the sample.}
\end{figure}

\begin{figure}
\label{edropbig}
\centerline{\psfig{file=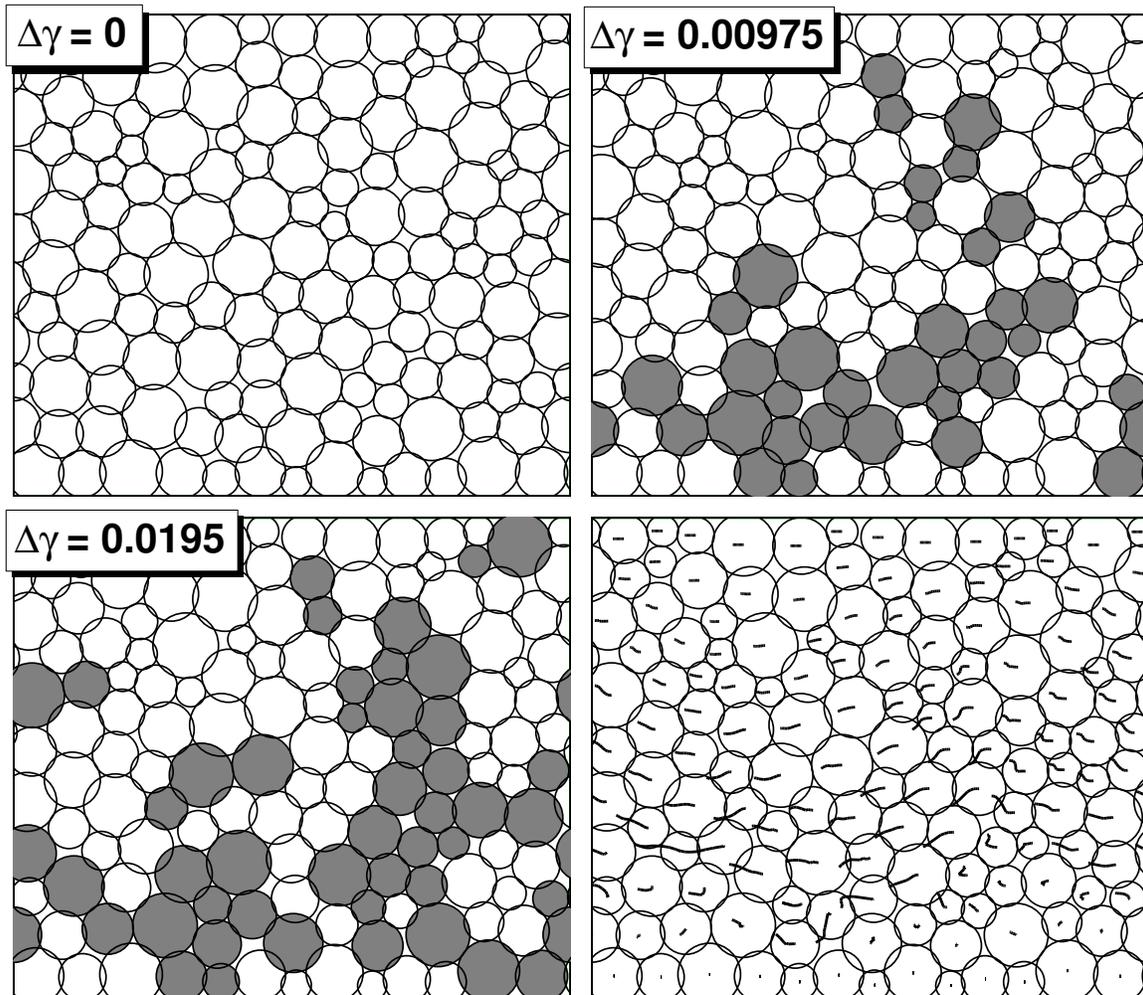,width=6in}}
\caption{Sequence of snapshots showing bubble rearrangements during a
large energy drop in a 144-bubble system at $\phi=1.0$ and
$\dot\gamma=10^{-3}$.  The magnitude of the drop ($\Delta E/E_{b}=13.18$)
and the fraction of bubbles that change overlapping neighbors
($N/N_{bub}=0.44$) both fall in the upper tails of the distributions.  The
first three frames show the configurations at the beginning, middle
and end of the energy drop; the gray bubbles have changed overlapping
neighbors since the start of the drop.  The fourth frame shows the
final configuration along with the tracks made by the centers of the
bubbles during the event.  We did not use the same scheme as in the
fourth frame of Fig.~2 to show the rearrangements because the bubble
motions were too extensive in this case.}
\end{figure}

While useful for building intuition, the distribution of energy drops
does not yield direct information about bubble rearrangements.
Therefore, we also measure the number $N$ of bubbles that experience a
change in overlapping neighbors during an energy drop.  We exclude
events in which two bubbles simply move apart or together; thus the
smallest event is $N=3$.  A typical sequence of configurations before, during,
and after an event is shown in the first three frames of Fig.~\ref{edropav}.
In this energy
drop the magnitude of the drop and the number of bubbles that change
neighbors are close to the average.  In the second and third frame of
the sequence, we have marked
the bubbles that changed neighbors since the beginning of the energy
drop (shown in the first frame).  As the system is strained, more
bubbles change neighbors.  For the particular energy drop chosen,
roughly one-sixth of the bubbles eventually change overlapping
neighbors.  The fourth frame shows the final configuration of
bubbles (colored gray) superimposed on the initial configuration at
the start of the energy drop (colored black).  Most of the bubble
motions that lead to this average-sized energy drop are rather subtle
shifts; there are no topological rearrangements.  A large energy
drop, from the tail of the distribution, is shown in
Fig.~\ref{edropbig}.  Again, the first three frames show the
configurations at the
beginning, middle and end of the drop, with the bubbles that change
overlapping neighbors marked in gray.  The fourth frame shows the
extensive rearrangements that occur from the beginning to the end of
the drop.  The configuration shown is the final one, and the short segments
are the tracks made by the centers of the bubbles during
the energy drop.

Typically, larger drops involve
larger numbers of bubbles.  Fig.~\ref{fig1}c depicts $N$ during each
energy drop in the same run as in
Fig.~\ref{fig1}a and b.  (Here, $N$ is normalized by the total number
of bubbles in the system, $N_{bub}$.)  The correlation between energy drops
and
the
number of bubbles involved
is shown by a scatter plot of these quantities in Fig.~\ref{scat} for
a 900-bubble system strained from 0 to 10.  We see that indeed there is a
strong correlation between these two measures of the size of an event.
Larger drops in energy involve larger numbers of bubbles and are
therefore spatially more extended.  The correlation is particularly
good at the large-event end.  There is more variability for midsize
and small events -- a large range of energy drops corresponds to the
same small number of rearranging bubbles, suggesting that typical
rearrangements involve only a few bubbles.

Besides counting statistics for energy drops and changes in number of
bubble overlaps, another direct measure of
bubble rearrangements is the rate of T1 events, i.e. of topology
changes of the first kind \cite{weaire84}.  For a perfectly dry
two-dimensional foam consisting of thin films, these are said to occur
when a bubble edge shrinks to zero, such that a common vertex is shared by
four bubbles, two moving apart and two moving together.  These events
were the only property used by Dennin and Knobler~\cite{dennin} to
characterize the response of their monolayer foam to shear because
they were unable to measure changes in the energy.  While the time at
which a T1 event occurs is well defined in a dry foam, it is somewhat
ambiguous for a wet foam because there can be an exchange of nearest
neighbors without a common point of contact.  Moreover, while the
number of bubbles involved in a T1 event is four by definition, large
clusters of bubbles can rearrange, with some of the interior bubbles
being involved in two or three T1 events simultaneously.  It is then
much harder to assign an exact time to a T1 event.

To make contact with the monolayer experiments, we may define T1
events within the bubble model as follows.  First we broaden the
definition of ``nearest neighbors'' to also include bubbles that do
not necessarily overlap, but that are nonetheless so close such that
$|\vec{r}_i - \vec{r}_j| < a(R_i + R_j)$, where $a>1$ is a suitably
chosen factor that may depend on $\phi$.  We then say that a T1 event
begins when two nearest neighbors move apart, and we say that it ends
when a new nearest neighbor pair intrudes between them; the time at
which the event occurs is taken as the midpoint in this sequence.
This definition is illustrated in the time sequence of a T1 event
shown in Fig.~\ref{T1}.  While the duration
of an actual T1 event in a dry foam is instantaneous, the duration
within the bubble model may vary greatly.  Furthermore, the midpoint
in the sequence does not necessarily coincide with the exact moment
the switching occurs.  In many instances it takes a long time after
two bubbles separate for the remaining pair to come into contact.  To
compare with our other measures of rearrangement, we depict in Fig.~\ref{fig1}d
the number of T1 events as a function of strain for the same run
as in Figs.~\ref{fig1}a, b and c.  There appears to be good correlation
between the largest energy drops and instances in which many T1 events
occur simultaneously.  However, there are many more T1 events than
energy drops.  This is because many T1 events can occur when a large
cluster of bubbles rearranges, and because our definition also includes
topology changes that cause an {\it increase} in the total elastic energy.

We can examine the consequences of our definition of a T1 event by
studying the distribution of the number of rearrangement events as a
function of their total duration in units of the strain.  This is done for
both energy drops
and T1 events, as shown in Fig.~\ref{durdist}.  The duration of an
energy drop is taken as the difference in strain between a decrease in the
elastic
energy and the next increase.  It is evident from the duration
distribution for energy drops, Fig \ref{durdist}a, that most energy
drops occur over a relatively short strain scale.  In units of time,
the longest events are comparable to a hundred times the
characteristic time scale in the problem ($\tau_d=1$ in our
simulations).  We find a good correlation between the number of
bubbles that change overlapping neighbors and the duration of the
event; the more bubbles involved in the event, the longer it lasts.
The distribution for T1
events, shown in Fig \ref{durdist}b, has a qualitatively similar shape,
exhibiting a slightly more rapid decrease for both fast and slow
events.  However, the scale on which T1 events occur is
an order of magnitude larger than the characteristic duration of the
energy drops.  By examining the bubble motions we see that the largest
energy drops are associated with many T1 events,
but the difference in strain scales makes it difficult to demonstrate an
exact correlation between the number of overlap changes and the
number of T1's.  In counting the T1 events, we include only events
that have a total strain duration of less than 2.  Fig.~\ref{durdist}b
shows that we have included all the T1 events for this
run.

\section{Simulation Results}

For a given system size, strain rate, dissipation mechanism and gas
fraction, we now collect statistics on the
following measures of bubble dynamics: (1) The probability
distribution $P(\Delta E)$ for energy drops of size $\Delta E$;
(2) The probability distribution $P(N)$
for the number of bubbles $N$ that change overlapping neighbors during
a energy drop event; and (3) The event rates for both energy drops and
T1 events, ${S}(T1)$ and ${S}(\Delta E)$, both
defined as the number of events per bubble per unit strain.

\subsection{System Size}

We first address the important issue of the finite
size of the simulation sample.  This is done for dry foams,
$\phi=1.0$, driven at a slow strain rate, $\dot \gamma=10^{-3}$.  The
results for four system sizes, $N_{bub}=36$, 144, 324 and 900, are shown in
Fig.\ref{size}.  In these runs, the systems were strained up to
80, 80, 31 and 10, respectively.
The top plot shows the energy drop distribution scaled by $E_b$, the
average energy per bubble.  It shows that energy drops
vary greatly in size over the course of a single
run.  The general features of this distribution have been reported
earlier \cite{durian2}.  There is a power-law region with an exponent
of -0.7 that extends over several decades in $\Delta E/E_b$, followed
by a sharp cutoff that occurs above a characteristic event size.  Such
a distribution has a well-defined average energy drop, which is near
the cutoff between 2$E_b$ and 3$E_b$ for the systems shown here.  The
slight deviation from power-law behavior for small $\Delta E$ was
absent in the earlier simulations \cite{durian2}, which did not
exclude two-bubble events, and which had a different roundoff error.  Also,
as seen earlier \cite{durian2}, the
two largest systems, with 324 and 900 bubbles, respectively,
have nearly identical
distributions.  This has two important implications; namely, that the
sharp cutoff of the power-law distribution is not a finite-size
effect, and that the system does not exhibit self-organized
criticality.

The presence of a characteristic energy-drop size can be corroborated by
examining the number of bubbles that participate in rearrangements for the
same set of
runs, which is given in the middle plot, Fig. \ref{size}b.  This quantity
has not been studied
previously
within the bubble model.  We plot the probability distribution
$P(N)$ of the number of bubbles $N$ that change overlapping neighbors during a
rearrangement.  The distribution decreases monotonically with a sharp
cutoff at the
large-event end.  This indicates that most of the rearrangements are
local and involve only a few bubbles.  Fig.~\ref{size}b shows that as the
system size
increases, the largest events represent a smaller fraction of the
total number of bubbles.  Indeed, the tail of the distribution extends to
smaller and
smaller values of $N/N_{bub}$ with no signs of saturation as the system
size $N_{bub}$ increases, indicating diminishing finite size effects.

We next look at the system-size dependence of event rates,
$S(T1)$ and ${S}(\Delta E)$, for the number of T1
events and energy drops per bubble per unit strain.  This is shown in
the bottom plot, Fig. \ref{size}c, for the same runs as in Figs.
\ref{size}a-b.  We find that ${S}(\Delta E)$ decreases very
slightly with increasing system size, but saturates for the largest
systems.  The results for ${S}(T1)$ show a stronger
system-size dependence, increasing slightly with $N_{bub}$.  This could be
due to the fact that bubbles on the top and bottom boundaries of the
system are fixed, which lowers the number of possible T1 events per
bubble.  As the system size grows, the boundary bubbles represent a
smaller fraction of the system so the event rate increases towards its
bulk value.

In short, all of our measurements at $\phi=1.0$ and $\dot
\gamma=10^{-3}$ indicate that the rearrangement events are localized
and that there is no self-organized criticality.  This agrees with
observations of rearrangements in both monolayer and bulk foams.

\subsection{Shear Rate Dependence}

Now that size effects have been ruled out for dry foams, we may
examine the influence of shearing the sample at different rates.
Experiments by Gopal and Durian\cite{gopal2} on three-dimensional
foams show a marked change in the character of the flow with
increasing shear rate.  At low shear rates, the flow is characterized
by intermittent, jerky rearrangement events occurring at a rate
proportional to the strain rate.  As the shear rate increases, so that
the inverse shear rate becomes comparable to the duration of a
rearrangement event, the flow becomes smoother and laminar, with all
the bubbles gradually rearranging all the time.  This was attributed
to a dominance of viscous forces over surface tension forces when the
strain rate exceeds the yield strain divided by the duration of a
rearrangement event.
In movies of our simulation runs, we also observe a crossover from
intermittent, jerky rearrangements to smooth laminar flow.  Similar
smoothing has also been seen in stress vs. strain at increasing shear
rates for the mean-field version of bubble dynamics \cite{durian2}.
This raises the question of how the statistics of rearrangement events
change with shear rate.  Specifically, how is the ``smoothing out''
of the flow reflected in the statistics at high
rates, and is there a quasistatic limit at low shear strain rates, in which
rearrangement behavior is independent of strain rate?  Earlier numerical
studies by
Bolton and Weaire\cite{bolton} were restricted, by construction, to
the quasistatic limit.  Okuzono and Kawasaki\cite{okuzono} examined
nonzero shear rates, but focused only on establishing the low
shear-rate limit.  Recently, Jiang and coworkers found a strong
dependence of the T1 event rate on shear rate \cite{jiang}.  They
found that the number of T1 events per bubble per strain,
${S}(T1)$, decreases sharply with strain rate with no evidence of a
quasi-static limit.

Our results for rearrangement behavior vs strain rate are collected in
Fig.  \ref{shear} for a 144-bubble system at $\phi=1.0$.  The top plot for
the probability distribution of energy drops indicates that there is
no gross change in $P(\Delta E)$ with shear rate, even though our
movies show a smoothing with less frequent energy drops.  However,
there is some suppression of small energy drops with an accompanying
increase at large energy drops, as reflected in a somewhat smaller
power-law exponent and larger cutoff at high values of $\Delta
E/E_{b}$.  It is not apparent from $P(\Delta E)$ vs $\Delta E/E_{b}$,
but we find that the average energy drop $\langle \Delta E \rangle$
and the average energy per bubble $E_{b}$ both increase with shear
rate, and that $\langle \Delta E \rangle$ increases more rapidly.
The reason why $E_{b}$ increases with shear rate is, of course, that
viscous forces become more important than elastic forces and lead to
increasing deformation (or in our model, overlaps) of bubbles.  The
net result is that there are fewer, relatively larger, rearrangements
at high strain rates.

The tendency that small events are suppressed with increasing shear
rates is also borne out by the distribution of the number of bubbles
that change neighbors during an energy drop, as shown in
Fig.~\ref{shear}b.  Note that unlike the previous curves, $P(N)$ is
plotted here on a linear scale.  Two systematic trends emerge with
increasing $\dot{\gamma}$: there are relatively fewer small events,
i.e. $P(N)$ decreases significantly at small $N/N_{bub}$, and the tail extends
to slightly higher $N/N_{bub}$.  For $\dot{\gamma} \geq 10^{-1}$ the
distribution is
fairly flat, suggesting that no one event size is dominant and there
are numerous large events of the order of the system size.  This
suggests that at this shear rate the system no longer relaxes stress
by intermittent rearrangements, but by continuous flow, as confirmed by our
movies of the runs \cite{movies}.  The trend
in $P(N)$ is seen in larger systems as well.  For the 900-bubble system we
also find that as the shear rate increases from $10^{-5}$ to
$10^{-3}$, the distribution flattens and extends to higher values of
$N$.  The average number of rearranging bonds increases
with shear rate, consistent with the picture of many bubbles in motion
as the system becomes more liquid-like.  We cannot, however, probe the
system at very high shear rates.  Data above a shear rate of about 1
cannot be trusted because of the nature of the model used.  At high
rates of strain the viscous term dominates and the elastic forces are
not strong enough to prevent clumping of bubbles.  This is actually an
artifact of the assumption that only overlapping bubbles interact
viscously; such clumping does not occur until much higher strain rates
in the mean-field version of dynamics.  Another reason why we do not
study shear rates higher than unity is because we do not allow bubble
breakup under flow (recall that $\dot\gamma$ is the capillary number).

The gradual smoothing with increasing shear rate is most apparent in
Fig \ref{shear}c, where we see that the event rates of
T1 events and energy drops both decrease
with increasing strain rate.  For the T1 events, the decrease is
slight, and is primarily due to the fact that the event duration
becomes even longer.  The decrease is more
dramatic for the energy drop events.  With increasing strain rate, the
average energy drop
increases and the rate of energy drops decreases.

Let us now re-examine the behavior of all quantities in
Fig.\ref{shear}, focusing on behavior at low shear strain rates.
Note that all quantities appear to approach a
reasonably well-defined ``quasistatic'' limit insensitive to the value
of $\dot\gamma$.  We thus have the following picture.  For small
$\dot{\gamma}$, the time between rearrangements is typically much
longer than the duration of a rearrangement, implying there is
adequate time for the system to relax stress.  As the shear rate
increases, bubbles are constantly in motion and cannot fully rearrange
into local-minimum-energy configurations.  Therefore, the viscous
interactions dominate, and the system flows like an ordinary liquid.

\subsection{Mean-Field vs Local Dissipation}

In the bubble model at higher strain rates, the behavior was seen to
depend on the form of dissipation: clumping for local dissipation,
Eq.~\ref{local}, as opposed to no clumping for mean-field dissipation,
Eq.~\ref{mf}.  In this section we will investigate whether dissipation
affects the low-strain-rate behavior as well.  If there truly exists a
quasi-static limit as $\dot{\gamma}\rightarrow 0$, as suggested by the
plots in the previous section, then the form of dissipation should
have no influence.  This need not occur, since once a rearrangement
starts it proceeds with finite speed according to dynamics set by a
competition between surface tension and dissipation forces.  For
example, it is conceivable that the mean-field dynamics might
discourage the mushrooming of a tiny shift in bubble position into a
large avalanche, whereas local dynamics might not.  Another important
issue is that differences in mean-field vs local dissipation could be
relevant to true physical differences between bulk foams and Langmuir
monolayers at an air/water interface.  For three-dimensional foams,
the shear is transmitted through the sample via bubble-bubble
interactions, so the dissipation might be better captured by the
local dissipation model.  In contrast, for two-dimensional Langmuir
monolayer foams the subphase imposes shear on the monolayers, and the
dissipation might therefore be closer to that calculated with the
mean-field model.

To investigate the influence of mean-field vs local dynamics, we can
simply compare avalanche statistics.  This is done in Fig. \ref{MFL}
for 144-bubble systems at four different area fractions, all sheared at
$\dot\gamma=10^{-3}$.  The top plot shows results for
the energy-drop distribution, $P(\Delta E)$, with solid/dashed curves
for local/mean-field dissipation respectively.  There is no
significant difference seen between the two choices of dissipative
dynamics.  This is also true of the spatial extent of the
rearrangements, as seen in the middle plot for the probability
distribution $P(N)$ of rearranging bubbles.  The bottom
plot for the rate of energy-drop and T1 events also shows little
significant difference between mean-field and local dynamics.  The
only distinction is a slightly greater rate of T1 events in the
mean-field case.  This reflects the difference in duration of T1
events within the two models; we find that T1 events tend to last
longer within the local dissipation model.  Since we do not count T1
events that last longer than a strain of 2, we count fewer events within
the local model than the mean-field version.  Thus, the differences in
$ S(T1)$ may simply be due to our method of counting T1
events.  Taken together, the three plots in Fig.~\ref{MFL} encourage
us to believe that the rearrangement dynamics predicted by the model
are robust against details of the dissipation.  They also provide
further evidence for the existence of a true quasi-static limit, where
the effect of strain rate is $\it{only}$ to set the rate of
rearrangements.

\subsection{Gas Area fraction}

Finally, we turn to the issue of how the elastic character of a foam
disappears with increasing liquid content, and possibility of critical
behavior at the melting transition.  The principal signature of the
melting, or rigidity-loss,
transition is that the shear modulus $G=\lim_{t\rightarrow\infty}
\sigma(t)/\gamma$ vanishes and the foam can no longer support a
nonzero shear stress without flowing.  In two-dimensional systems,
this happens at a critical gas fraction corresponding to that of
randomly packed disks, $\phi_c \approx 0.84$.  This has been seen in
several different simulations, where the gas fraction was tuned to
within 0.05 of the transition\cite{bolton0,bolton,hutzler} and where
it was tuned through, and even below, the
transition\cite{durian1,durian2}.  Other signatures of melting are
that the osmotic pressure vanishes as a
power-law\cite{durian1,durian2,lacasse} the coordination
number decreases towards about
4 as a power-law\cite{bolton0,bolton,hutzler,durian1,durian2,weaire}, and
that the time
scale for stress relaxation following an applied step-strain appears to
diverge \cite{durian1,durian2}.
Here we look for signs of melting in the
statistics of avalanches during slow, quasi-static flow.  Within our
model, an increase in liquid content causes a decrease in the average
overlap between neighboring bubbles.  This in turn produces a decrease
in the average elastic energy of the system, $E_{b}$ and sets the scale
for the average energy drop
$\langle\Delta E\rangle$ per rearrangement.  It therefore should also
decrease at lower gas fractions $\phi$.

The energy drop and size statistics of rearrangement events for
increasingly wet foams were shown already in Fig.~\ref{MFL}, but were
discussed only in the context of mean-field vs local dissipative
dynamics.  A clear trend emerges when we examine the $\phi$ dependence
specifically.  In the top plot Fig.~\ref{MFL}a for $P(\Delta E)$, we
see that the power-law behavior for small events does not change, but
that the exponential cut-off moves towards larger values of $\Delta
E/E_b$ as $\phi\rightarrow\phi_{c}$.  Though both $\langle\Delta
E\rangle$ and $E_{b}$ decrease towards zero, the latter evidently
vanishes more rapidly.  This results in a broader distribution of
event sizes near the melting transition; as the system becomes more
liquid, large events are more prevalent.
The probability distribution $P(N)$ for the
numbers of bubbles involved in rearrangement events is shown in
Fig. \ref{MFL}b.  It displays similar trends as a function of
$\phi$, but not as pronounced as in $P(\Delta E)$.
Namely, the power law for small $N$ is unaffected by $\phi$, but the
exponential cut-off moves towards slightly larger events as
$\phi\rightarrow\phi_{c}$.  Thus, although the scale of energy drops
increases dramatically, the number of broken bonds only increases
marginally.  Note,
however, that the largest events include almost all the bubbles in the
system; thus, the relatively weak dependence of $P(N)$ on $\phi$ could
be a finite-size effect in these $N_{bub}=144$ systems, as we will
show below.

The behavior of ${S}$, the number of energy drops and T1
events per bubble per strain, is shown in Fig.\ref{MFL}c.  As the
system becomes wetter, there is no noticeable change in the event rate
${S}(\Delta E)$ for energy drops.  In contrast, if our
definition of nearest neighbors only includes overlapping bubbles, we find
that ${S}(T1)$ decreases as $\phi$
decreases.  This runs counter to expectations--bubbles in a wet foam
should have more freedom to move and rearrange because the energy
barrier between rearrangements is lower and the yield strain is
smaller.  The apparent drop arises because the
bubble coordination number is much higher in a dry foam (roughly 6) than in a
wet foam (roughly 4).  As a result there are more overlapping neighbors for
each bubble
in a dry foam, and more possibilities for the occurrence of T1 events.
In the wet foam, however, there are many T1 events that do not satisfy
the stringent starting or ending configurations because neighboring
bubbles do not overlap.  It is therefore appropriate in wet foams to
modify the criterion for neighbors to $|{\bf r}_i - {\bf r}_j| < a(R_i
+ R_j)$, where the proximity coefficient $a$ is taken as $1/\phi$.
When T1's are computed with this definition, we find no significant
dependence on area fraction.

The fact that the power-law region of the energy drop distribution is
more extended at lower area fractions suggests the possibility of a
critical point as the close-packing density, $\phi_{c}$, is approached
from above.  This would imply a pure power-law distribution $P(\Delta
E)$ for the energy drops at $\phi_{c}$, which would presumably be
accompanied by a growing correlation length, as well as the growing
relaxation time observed previously in Refs.~\cite{durian1,durian2}.
Note, however, that the distribution of the number of bubbles involved
in a rearrangement, $P(N)$, does not depend very strongly on $\phi$
for the 144-bubble systems of Fig. \ref{MFL}; furthermore, the
cut-off to power-law behavior is always present, no matter how closely
$\phi_{c}$ is
approached.  This raises the question of whether finite system size
effects are more important at values of $\phi$ near $\phi_{c}$ (recall
from Fig.~\ref{size} that there were no significant system size
effects near $\phi=1$).  To examine this, we have plotted the
dependence of $P(\Delta E)$, $P(N)$ and $ S$ on system size
in Fig.~\ref{size:85}.  We indeed find a strong system size dependence
in $P(\Delta E)$ at $\phi=0.85$ just above the melting transition, with
no saturation at the largest size studied (900 bubbles).  This is consistent
with the existence of a long correlation length.

The distribution of the number of bubbles per
energy drop, $P(N)$ also shows signs of
criticality.  Recall from Fig.~\ref{size}b that at $\phi=1$, the tail of
$P(N)$ was cut off at smaller and smaller values of $N/N_{bub}$ with
increasing
system size at $\phi=1$.  This was consistent with a short correlation
length, characteristic of localized rearrangement events.  At
$\phi=0.85$, the behavior with increasing $N_{bub}$ is quite
different, as shown in Fig.~\ref{size:85}b. The distribution falls
off slightly more rapidly with $N/N_{bub}$ at larger system sizes
(probably because $\phi=0.85$ still lies above $\phi_{c}$), but the largest
events in the system still involve the same fraction $N/N_{bub}
\approx 0.75$ of bubbles, indicating a correlation length that is
comparable to the largest system size studied (30 bubble diameters
across).

The event rates for energy drops and T1 events for the
different system sizes at $\phi=0.85$ are shown in Fig.
\ref{size:85}c.  The behavior is not markedly different from that
found for the drier foam.  Recall, however, that we have adjusted our
definition of a T1 event by changing the proximity coefficient $a$
with area fraction, so little can be expected to be learned from this
measure.

\section{Discussion}

\begin{figure}
\label{scat}
\centerline{\psfig{file=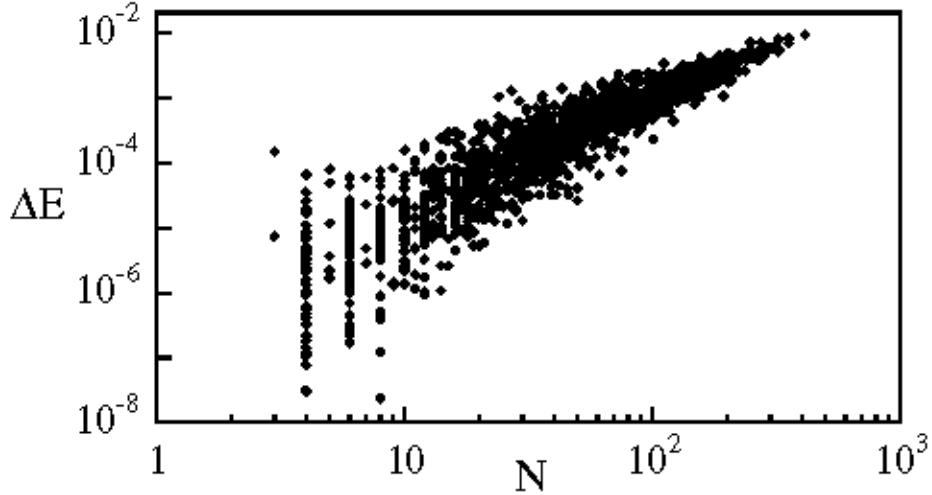,width=5in}}
\caption{The size of energy drops as a function of the number of
bubbles that concurrently change overlapping neighbors during the
energy drop,
for a 900-bubble system at $\phi = 1.0$ driven at $\dot{\gamma}=
10^{-3}$.  This indicates that the fraction of bubbles that change
overlapping neighbors during an energy drop
increases with the size of the energy drop.}
\end{figure}

\begin{figure}
\label{T1}
\centerline{\psfig{file=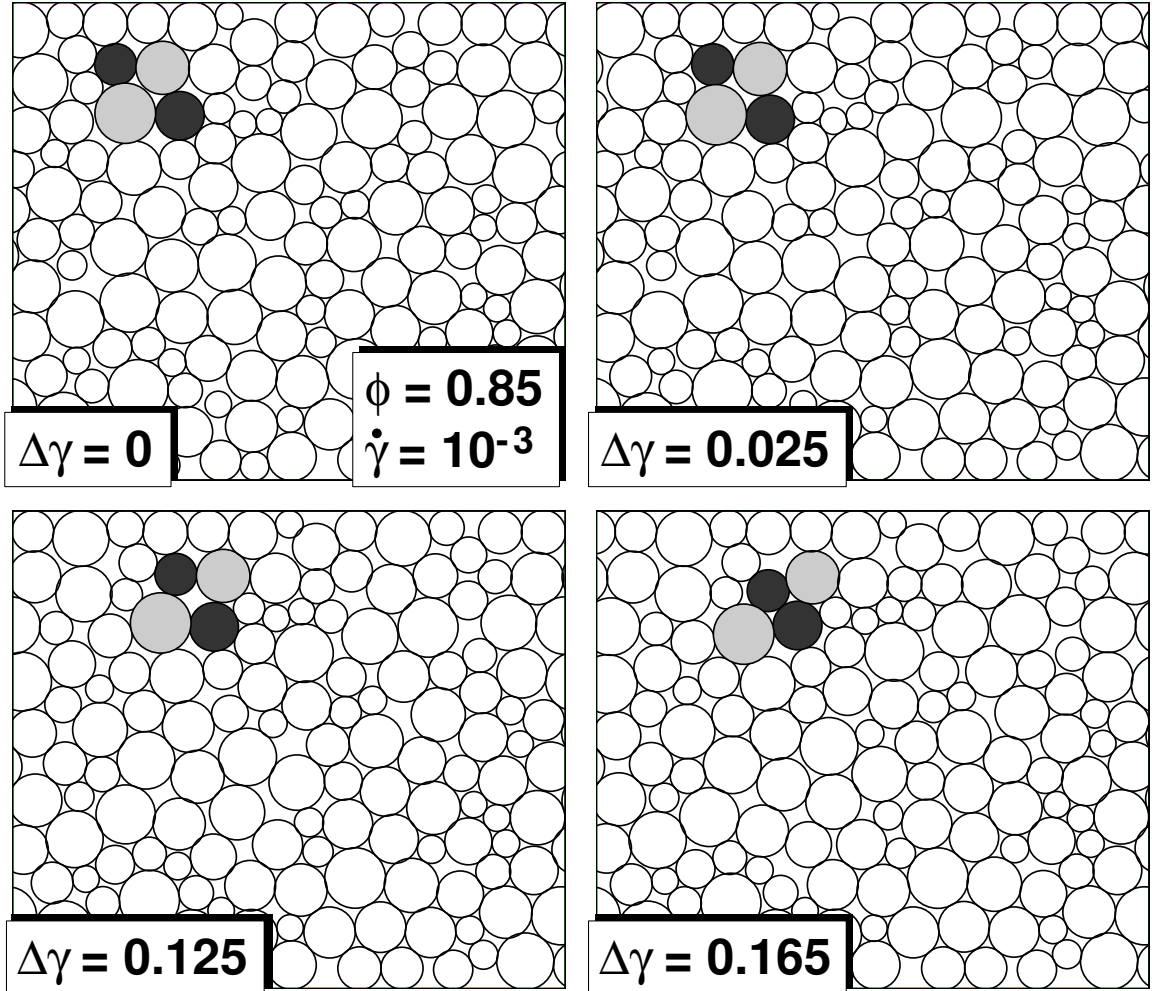,width=6in}}
\caption{A sequence of snapshots showing a T1 event in a wet
foam at $\phi=0.85$ as the system is strained at $\dot{\gamma}=
10^{-3}$.  During the event, the black
pair of bubbles moves together and the gray pair moves apart.}
\end{figure}

\begin{figure}
\label{durdist}
\centerline{\psfig{file=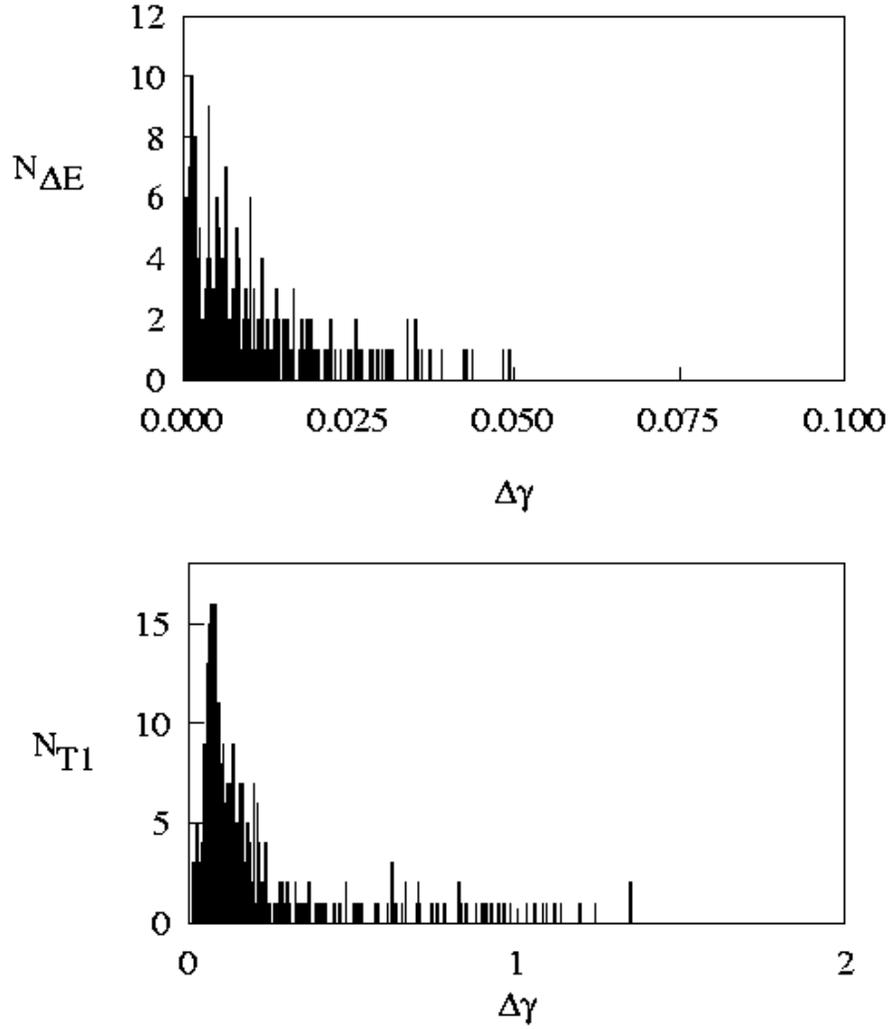,width=5in}}
\caption{The probability distribution for the duration of (a)
energy-drop rearrangement events and (b) T1 events, for a 144-bubble system
$\phi=1$ driven at $\dot{\gamma}=10^{-3}$.  Note that the typical
duration of T1 events is significantly longer than that of energy drops.}
\end{figure}

\begin{figure}
\label{size}
\centerline{\psfig{file=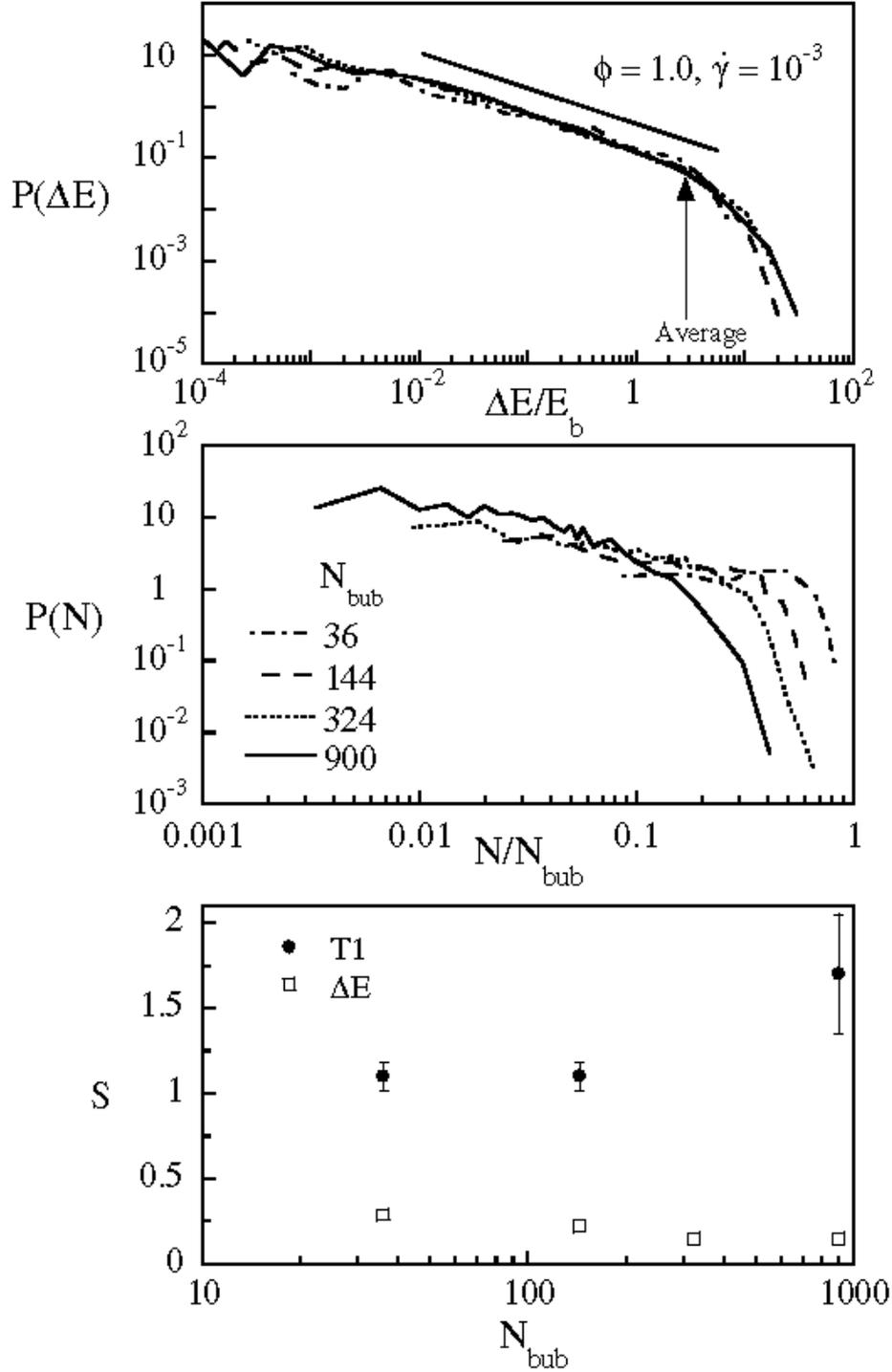,width=5in}}
\caption{ Effect of system size at $\phi=1.0$ and
$\dot{\gamma}=10^{-3}$.  (a) Probability density distribution of
energy drops $\Delta E$ scaled by $E_b$, the average energy per bubble
for each run.  There is a power-law region (the straight line has a
slope of -0.7)
followed by a sharp cutoff.  The cutoff depends only weakly on
the system size and converges for the larger systems.  (b)
Probability distribution of the number of bubbles that change
overlapping neighbors during a rearrangement.  The tails
of the distribution extend to smaller fractions of the total number of
bubbles in the system as the system size
increases, showing that the events are spatially localized.  (c) Event rate
for T1 events (solid circles) and energy
drops (open squares).  Error bars in this and subsequent figures represent
the variations found in independent runs (at least three, with the
exception of the 900-bubble system for which only two runs were carried
out) with the same initial conditions.  Where no error bar is indicated the
variation is smaller than the sixe of the symbol. The number of energy
drops per bubble decreases
as the system size increases, reaching the same value for the
324-bubble
and 900-bubble systems.  There are, however, more T1 rearrangement events
per bubble at the larger system size.}
\end{figure}

\begin{figure}
\label{shear}
\centerline{\psfig{file=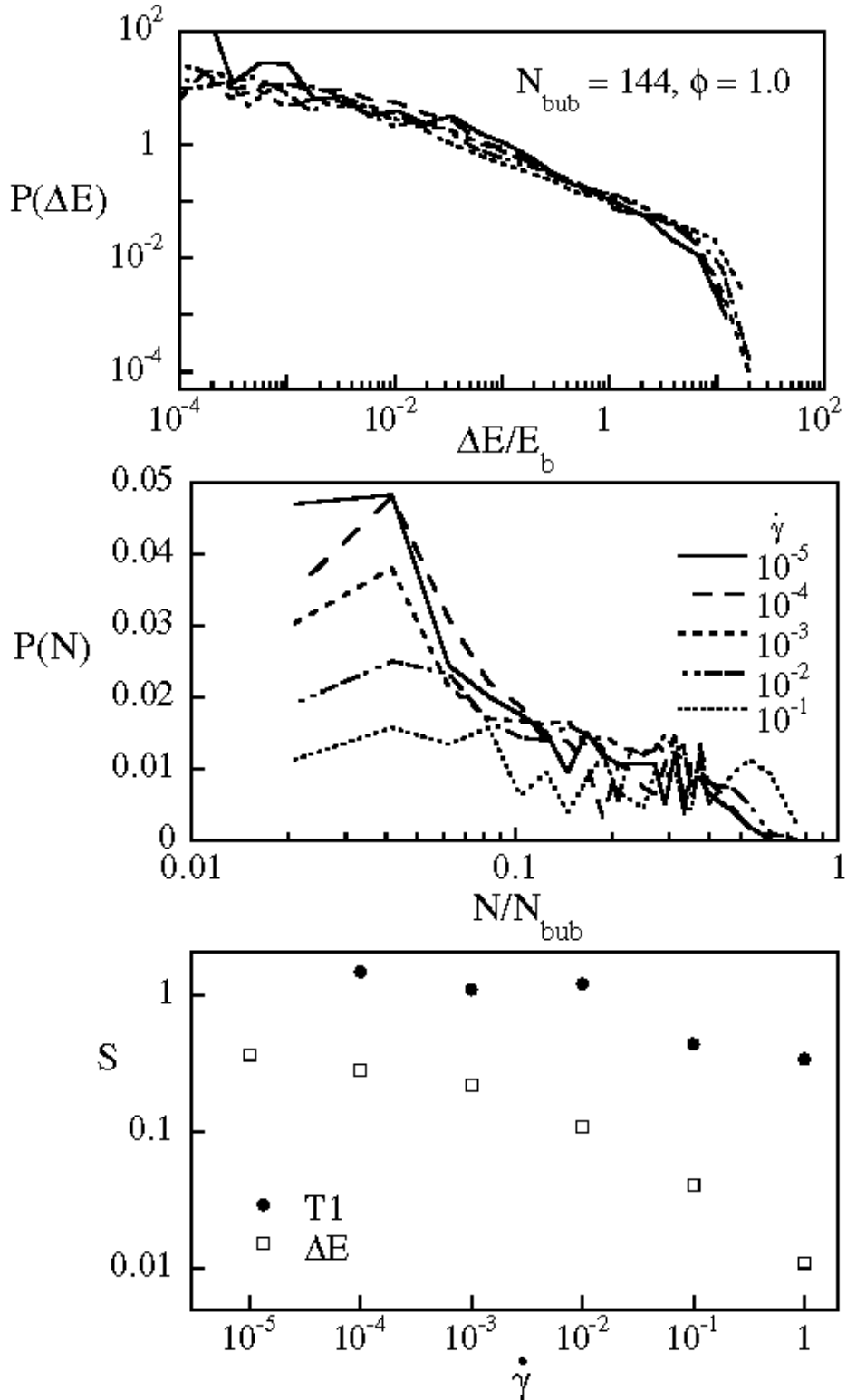,width=5in}}
\caption{Effect of shear rate for a 144-bubble system at $\phi= 1.0$.  (a)
There is no systematic change in the power-law region of the
probability distribution of energy drops.  The cutoff moves towards larger
event sizes as $\dot{\gamma}$ increases.  (b) A stronger trend is
apparent in the probability distribution of rearranging bubbles.  As
$\dot{\gamma}$ increases, the distribution flattens.  For the
highest rate,
$\dot{\gamma}=0.1$, the distribution is fairly flat, suggesting that no
one event size is dominant and the largest events are of the order of
the system size.  (c)  Both the event rates for T1 events and
energy drops decrease as the system is sheared faster.
The T1 event rates at $\dot{\gamma}=10^{-3}$ and $10^{-2}$ are the
same within error.  Note that a well-defined quasi-static limit is
approached as $\dot\gamma\rightarrow0$.}
\end{figure}

\begin{figure}
\label{MFL}
\centerline{\psfig{file=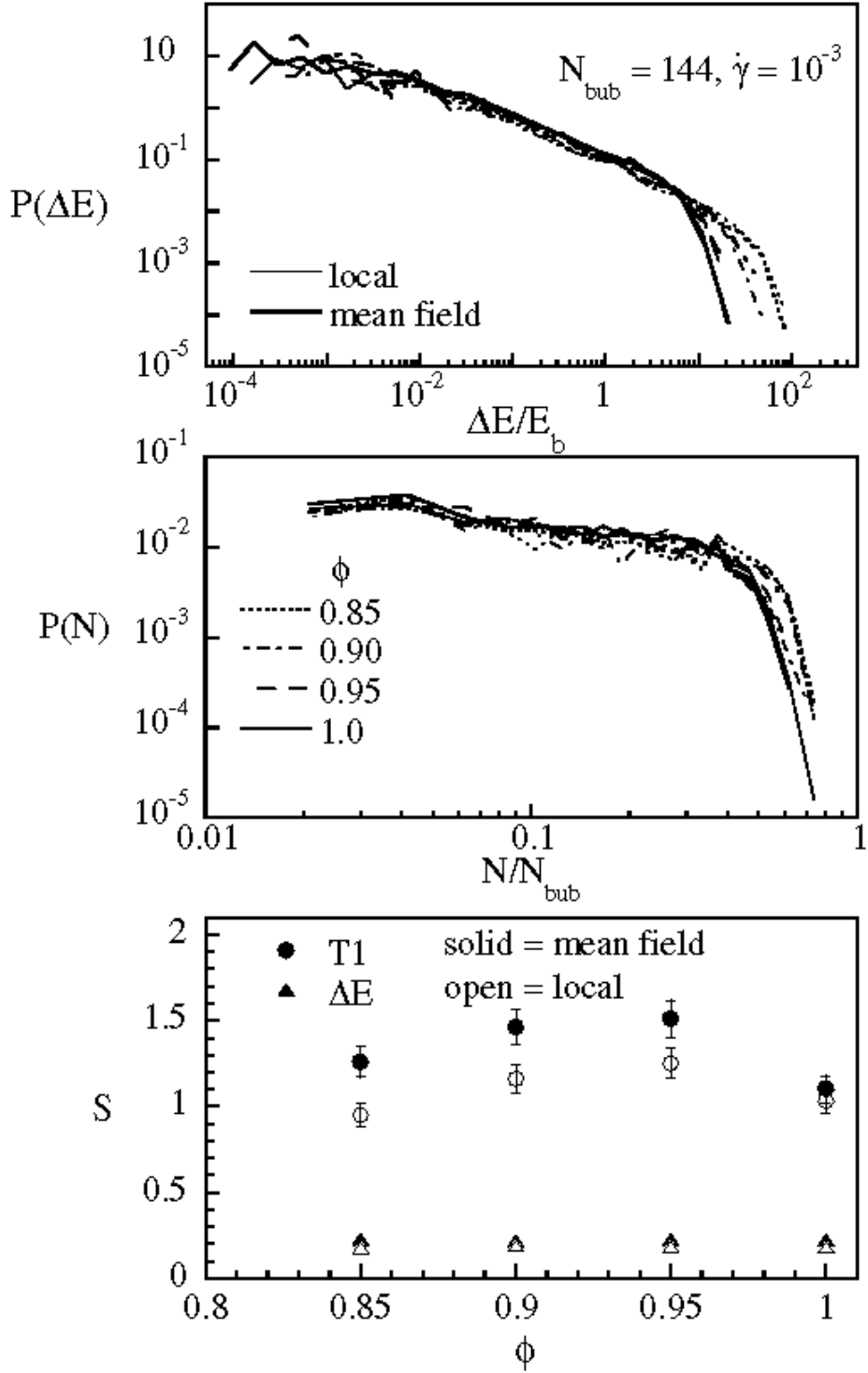,width=5in}}
\caption{Effect of gas area fraction and the form of viscous
dissipation for a 144-bubble system sheared at $\dot{\gamma}=10^{-3}$.
The probability distribution of both (a) energy drops, and (b) number
of bubbles changing overlapping neighbors during an event, are given
at four area fractions: $\phi=1.0$, 0.95, 0.90, and 0.85.  Heavy and
light curves are for mean-field and local versions of dissipative
dynamics, respectively.  Note that the dynamics do not influence the
behavior but that the events become larger as the gas fraction
approaches the melting point, $\phi_{c}\approx 0.84$.  Part (c) shows
the event rates for T1 events and energy drops; these are insensitive
to both gas area fraction and type of dynamics.}
\end{figure}

\begin{figure}
\label{size:85}
\centerline{\psfig{file=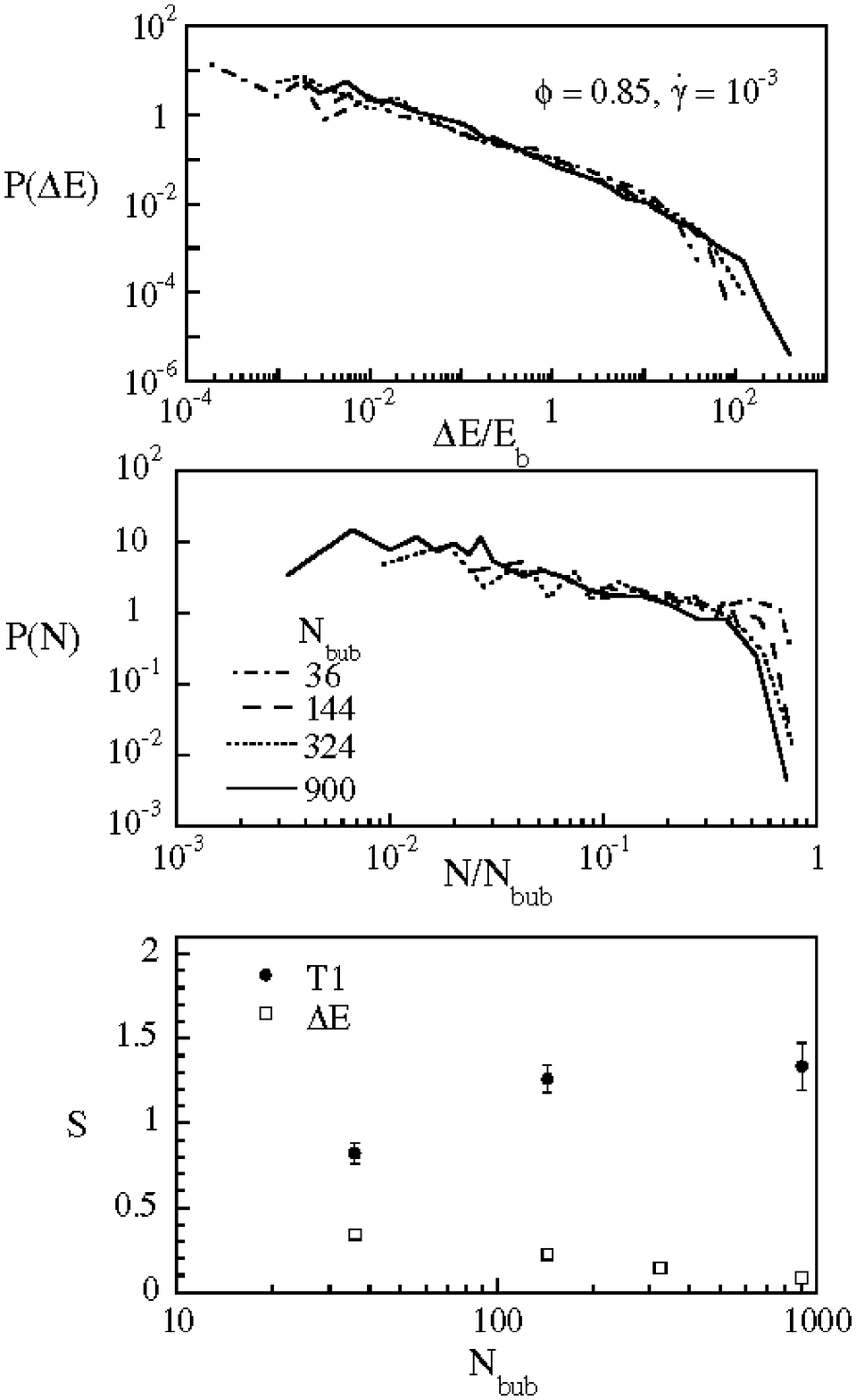,width=5in}}
\caption{ Effect of system size at $\phi=0.85$ and
$\dot{\gamma}=10^{-3}$.  (a) There is no change in the power-law
region of the probability density distribution compared with Fig.~6a.
However, the cutoff increases and there is no convergence
for the largest system sizes.  (b)  Even at the largest system sizes,
the largest events involve a significant fraction of the bubbles in
the system, indicating that the events are much more spatially
extended than at $\phi=1$.  (c) As in Fig.~7c, the number of T1 events
and energy drops show the opposite trend.  The event rates for energy
drops indicate that finite size effects are more pronounced at this
lower area fraction.  Unlike the saturation seen in Fig.~7c, the
event rate continues to drop as the system size increases.}
\end{figure}

We have reported the results of several different measures of
rearrangement event dynamics in a sheared foam.  A comparison of the
probability distribution of energy drops $P(\Delta E)$ with the
probabilty distribution of bubbles changing neighbors $P(N)$ shows
that the size of an energy drop correlates well with the number of
bubbles involved in a rearrangement (see Fig.~\ref{scat}).  This is
valuable because the energy drop-distribution has been widely studied
theoretically, but is very
difficult to measure experimentally.  The number of bubbles involved
in rearrangements, however, can be probed with multiple light
scattering techniques on three-dimensional foams\cite{gopal} and by
direct visualization of two-dimensional foams\cite{dennin}.  A study
of the rate of occurrence of topological changes (T1 events) provides
a further link to experiments.

In general, our results agree with experiments on
three-dimensional and two-dimensional foams.  Despite its simplicity,
the bubble model appears to capture the main qualitative features of a
sheared foam remarkably well.  For example, we find that the size of
rearrangement events is typically small at
low shear rates and at area fractions not too close to $\phi_{c}$.
This is in accord with experiments of Gopal and Durian\cite{gopal},
and Dennin and Knobler\cite{dennin}, as well as simulation results of
Bolton and Weaire\cite{bolton} and Jiang and coworkers\cite{jiang}.
Our results do not agree with those of Okuzono and
Kawasaki\cite{okuzono}, however, who found power-law distributions of
rearrangement events at $\phi=1$ in two dimensions.

The largest discrepancies between our results and those of others lie
in the statistics of T1 events.  We find that the number of T1 events
per bubble per unit strain is of order unity and is generally
insensitive to shear rate and gas area fraction.  Kawasaki et al.
\cite {kawasaki} found similar results: ${S}(T1) = 0.5$ and
no dependence on shear
rate.  In the Potts-model simulations \cite
{jiang}, however, ${S}(T1)$ is unity at $\dot{\gamma}=10^{- 3}$ but
falls to about 0.01
at $\dot{\gamma}=10^{- 1}$.

The monolayer experiments \cite{dennin} yielded values of $ S(T1)
\approx 0.15$, nearly an order of magnitude lower than predicted by
our simulations.  Durian \cite{durian2} reported a number of rearrangement
events
per bubble per unit strain for simulations of a 900-bubble system at
$\dot{\gamma}=10^{- 5}$ that was comparable to the monolayer result,
but he measured the number of energy drops per bubble per unit
strain, $S(\Delta E)$, not the T1 event rate, $S(T1)$.  Note that our
energy-drop event rate, $S(\Delta E)$, agrees well with Durian's
earlier result.

One might guess that the discrepancy between our measurement of
$ S(T1)$ and that of the monolayer experiment might lie in the
method of analysis used to count T1 events.  Unlike the simulations,
in which the number of T1 events can be computed from an analysis of
bubble positions as a function of time, the number of T1's in the
monolayer studies was determined by repeated viewing of videotapes of
the experiments and counting of the events as the foam cells reach
their midpoint configuration.  It seemed possible, then, that the
difference between the simulation and the experiment was the result of
a systematic undercounting of the number of the events.  To check this
possibility, the number of T1's in a simulation run was determined by
observations of the animated bubble motions.  The number of events
missed in this unautomated counting was only 2\% of the total.

We believe that the origin of the discrepancy between the T1 event
rates in the simulation and the monolayer experiment lies in the yield
strain.  While the yield strain in the model system is less than 0.2,
which is consistent with that measured in three-dimensional foams,
that in the monolayer foams is closer to unity.  Bubbles in monolayer
foams can therefore sustain very large deformations without inducing
rearrangements.  The T1 event rate should be inversely proportional to
the yield strain.  Thus, the ratio of $ S(T1)$ in the simulation
to $ S(T1)$ in the experiment should equal the ratio
of the yield strain in the experiment to the yield strain in the
simulation.  This is exactly what we find.

One of our main results is that a quasistatic limit exists within
the bubble model.  We find that the statistics of rearrangement events are
independent of shear rate at low shear rates.  This agrees with the monolayer
experiments\cite{dennin}, which measured T1 event rates at two different shear
rates, $\dot
\gamma=0.003 s^{-1}$ and 0.11$s^{-1}$.  Dennin and Knobler found no
noticeable difference in the T1 event rate, despite the fact that the shear
rates
studied differ by a factor of thirty.  In addition, Gopal and Durian
found that the number of rearrangement events per bubble per second in a
three-dimensional foam is given by the event rate in the absence
of shear plus a term proportional
to the shear rate.  In their case, the event rate was nonzero in the
absence of shear because of coarsening; we have neglected this effect
in our simulations.  However, we do find that the rearrangement event
rate per unit time is simply proportional to the shear rate at low shear
rates.
Thus, experimental results in both two and three dimensions
contradict the simulation results of Jiang, et
al.\cite{jiang}, which find no quasistatic limit, but agree with our
findings.

The form of dissipation used in the
bubble model is a simple dynamic friction, which does not capture the
hydrodynamics of fluid flow in the plateau borders and films in a
realistic way.  However, our results suggest that we may still be
capturing the correct behavior at low shear rates.  We find that the
rearrangement event statistics are the same whether we use mean-field
or local dissipation at low shear rates.
This suggests that the statistics are determined by elastic effects rather
than viscous ones at low shear rates, and that the behavior in that limit
should be independent of the form of viscous dissipation used.

Finally, our results as a function of gas area fraction imply that
there may be a critical point at the melting transition, as the area
fraction approaches the random close-packing fraction from above.
Previous studies showed that both the shear modulus and yield stress vanish
as power laws at the melting
transition\cite{bolton,durian1}, and that the stress
relaxation time appears to diverge\cite{durian1}.  Here, we have shown by
finite-size studies that there
is also a correlation length, characterizing the size of rearrangements,
which grows as one approaches the melting transition.  We also find that the
distribution of energy drops appears to approach a pure power law in that
limit.

The existence of a critical point at the melting transition remains
to be tested experimentally.  The vanishing of the shear modulus and
osmotic pressure at the transition has been measured by Mason and
Weitz\cite{mason} for monodisperse, disordered emulsions, and by Saint-Jalmes
and Durian for polydisperse gas-liquid foams\cite{arnaud}.  However,
these small-amplitude-strain rheological measurements could not test
whether there is a
diverging length scale for rearrangements in a steadily sheared system
at the melting transition.  On the other hand, Gopal and
Durian\cite{gopal} have measured the size of rearrangement events in
a gas-liquid foam, but only at packing fractions well above the melting
transition.  At lower packing fractions close to the melting
transition, the liquid drains too quickly from the foam due to gravity
to permit such measurements.  Experiments under microgravity
conditions should be able to resolve whether the melting transition is
indeed a critical point.

\acknowledgements We thank Narayanan Menon and Ian K. Ono for many
helpful discussions, and we thank Michael Dennin for performing the
visual analysis of the number of T1 events.  This work was supported
by the National Science Foundation through grants CHE-9624090 (AJL),
CHE-9708472 (CMK), and DMR-9623567 (DJD), as well as by NASA through
grant NAG3-1419 (DJD).

\end{multicols}

\begin{references}

\bibitem{prudhommebook} R. K. Prud'homme and S. A. Khan, eds., {\it
Foams: Theory, Measurement, and Application}.  Surfactant Science
Series {\bf 57}, (Marcel Dekker, NY, 1996).

\bibitem{djddaw} D. J. Durian and D. A. Weitz,
"Foams," in Kirk-Othmer Encyclopedia of Chemical Technology, 4 ed.,
edited by J.I. Kroschwitz (Wiley, New York, 1994), Vol. 11, pp. 783-805.

\bibitem{weaire84} D. Weaire and N. Rivier, Contemp.  Phys.  {\bf 25},
55 (1984).

\bibitem{kranikrev} A. M. Kraynik, Ann.  Rev.  Fluid Mech.  {\bf 20},
325 (1988).

\bibitem{gopal} A. D. Gopal and D. J. Durian, Phys.  Rev.  Lett.  {\bf
75}, 2610 (1995).

\bibitem{gopal2} A. D. Gopal and D. J. Durian, to appear in J. Coll.
I Sci.  (1999).

\bibitem{dennin} M. Dennin and C. M. Knobler, Phys.  Rev.  Lett.  {\bf
78}, 2485 (1997).

\bibitem{okuzono} T. Okuzono and K. Kawasaki, Phys.  Rev.  E {\bf 51},
1246 (1995).

\bibitem{bolton0} F. Bolton and D. Weaire, Phys.  Rev.  Lett.  {\bf
65} 3449 (1990).

\bibitem{bolton} F. Bolton and D. Weaire, Phil.  Mag.  B {\bf 65}, 473
(1992).

\bibitem{hutzler} S. Hutzler, D. Weaire and F. Bolton, Phil.  Mag.  B
{\bf 71}, 277 (1995).

\bibitem{durian1} D. J. Durian, Phys.  Rev.  Lett.  {\bf 75}, 4780
(1995).

\bibitem{durian2} D. J. Durian, Phys.  Rev.  E {\bf 55}, 1739 (1997).

\bibitem{jiang} Y. Jiang, P. J. Swart, A. Saxena, M. Asipauskas, and
J. A. Glazier, Phys.  Rev.  E, preprint (1998), cond-mat/9902111.

\bibitem{pozrikidis} X. F. Li, H. Zhou and C. Pozrikidis, J. Fluid
Mech.  {\bf 286}, 379 (1995).

\bibitem{lacasse} The breakdown of pairwise additivity is discussed
in: M.  D. Lacasse, G. S. Grest, D. Levine, T. G. Mason and
D. A. Weitz, Phys.  Rev.  Lett.  {\bf 76}, 3448 (1996),
mtrl-th/9603006; M. D. Lacasse, G. S. Grest and D. Levine, Phys.  Rev.
E {\bf 54}, 5436 (1996), mtrl-th/9603005.

\bibitem{libs} DDRIV3 is available from http://gams.nist.gov.
SPARSKIT2 was obtained from ftp://ftp.cs.umn.edu.

\bibitem{glazierweaire} J. A. Glazier and D. Weaire, J. Phys.:
Condens.  Matter {\bf 4}, 1867 (1992).

\bibitem{stavansrev} J. Stavans, Rep. Prog. Phys. {\bf 56}, 733
(1993).

\bibitem{movies} Quicktime movies of the bubble model simulation for fast and
slow shear may be viewed at http://math.nist.gov/mcsd/Staff/SLanger/foam.

\bibitem{kawasaki} K. Kawasaki, T. Okuzono, T. Kawakatsu and T. Nagai
in Proceedings of the International Workshop of Physics of Pattern
Formation, ed.  S. Kai (World Scientific, Singapore, 1992).

\bibitem{weaire} D. Weaire, S. Hutzler, Physica A {\bf 257}, 264
(1998).

\bibitem{mason} T. G. Mason, J. Bibette and D. A. Weitz, Phys. Rev.
Lett. {\bf 75}, 2051 (1995).

\bibitem{arnaud} A. Saint-Jalmes and D. J. Durian, preprint.

\end{references}
\end{document}